\documentstyle[12pt]{article}

\newlength{\extraspace}
\newlength{\extraspaces}
\newcommand{\be}{\begin{equation}
\addtolength{\abovedisplayskip}{\extraspaces}
\addtolength{\belowdisplayskip}{\extraspaces}
\addtolength{\abovedisplayshortskip}{\extraspace}
\addtolength{\belowdisplayshortskip}{\extraspace}}
\newcommand{\ee}{\end{equation}}
\newcommand{\bear}{\begin{eqnarray}
\addtolength{\abovedisplayskip}{\extraspaces}
\addtolength{\belowdisplayskip}{\extraspaces}
\addtolength{\abovedisplayshortskip}{\extraspace}
\addtolength{\belowdisplayshortskip}{\extraspace}}
\newcommand{\eear}{\end{eqnarray}}
\newcommand{\ba}{\begin{array}}
\newcommand{\ea}{\end{array}}
\renewcommand{\thefootnote}{\fnsymbol{footnote}}
\newcommand{\eq}[1]{\mbox{eq.~(\ref{#1})}}
\def\simlt{\mathrel{\lower2.5pt\vbox{\lineskip=0pt\baselineskip=0pt
           \hbox{$<$}\hbox{$\sim$}}}}
\def\simgt{\mathrel{\lower2.5pt\vbox{\lineskip=0pt\baselineskip=0pt
           \hbox{$>$}\hbox{$\sim$}}}}
\catcode`@=11
\newcount\@tempcntc
\def\@citex[#1]#2{\if@filesw\immediate\write\@auxout{\string\citation{#2}}\fi
  \@tempcnta\z@\@tempcntb\m@ne\def\@citea{}\@cite{\@for\@citeb:=#2\do
    {\@ifundefined
       {b@\@citeb}{\@citeo\@tempcntb\m@ne\@citea\def\@citea{,}{\bf ?}\@warning
       {Citation `\@citeb' on page \thepage \space undefined}}%
    {\setbox\z@\hbox{\global\@tempcntc0\csname b@\@citeb\endcsname\relax}%
     \ifnum\@tempcntc=\z@ \@citeo\@tempcntb\m@ne
       \@citea\def\@citea{,}\hbox{\csname b@\@citeb\endcsname}%
     \else
      \advance\@tempcntb\@ne
      \ifnum\@tempcntb=\@tempcntc
      \else\advance\@tempcntb\m@ne\@citeo
      \@tempcnta\@tempcntc\@tempcntb\@tempcntc\fi\fi}}\@citeo}{#1}}
\def\@citeo{\ifnum\@tempcnta>\@tempcntb\else\@citea\def\@citea{,}%
  \ifnum\@tempcnta=\@tempcntb\the\@tempcnta\else
   {\advance\@tempcnta\@ne\ifnum\@tempcnta=\@tempcntb \else \def\@citea{--}\fi
    \advance\@tempcnta\m@ne\the\@tempcnta\@citea\the\@tempcntb}\fi\fi}
\catcode`@=12

\newcommand{\nh}{\rm nonstandard Higgs boson}
\newcommand{\sm}{\rm Standard Model\ }

\newcommand{\gev}{\mbox{\rm GeV}}
\newcommand{\La}{\Lambda}

\newcommand{\lr}{\mbox{$SU(2)_L \times SU(2)_R$}}

\newcommand{\trace}{\mbox{\rm Tr}}
\newcommand{\pauli}{\mbox{$\vec{\tau}$}}
\newcommand{\vw}{\mbox{$\vec{w}$}}

\newcommand{\pdmu}{\mbox{$\partial_{\mu}$}}
\newcommand{\PDmu}{\mbox{$\partial^{\mu}$}}

\newcommand{\cL}{{\cal L}} 

\newcommand{\np}{Nucl.\ Phys.\ {\bf B}}
\newcommand{\pr}{Phys.\ Rev.\ }
\newcommand{\prd}{Phys.\ Rev.\ {\bf D}}

\newcommand{\prl}{Phys.\ Rev.\ Lett.\ }
\newcommand{\pl}{Phys.\ Lett.\ {\bf B}}

\newcommand{\rmp}{Rev.\ Mod.\ Phys.\ }

\begin{document}

\begin{titlepage}
 \begin{flushright}BUHEP 95-3\\ hep-ph/9502267\\ February 9, 1995
  \end{flushright}
\vspace{24mm}

\begin{center}
{\LARGE The profile of a nonstandard Higgs boson\\ at the LHC}

\vspace{5mm}
Dimitris Kominis\footnote{e-mail address: kominis@budoe.bu.edu}
\ \  and \ \
Vassilis Koulovassilopoulos\footnote{e-mail address:
 vk@budoe.bu.edu}\\*[3.5mm]

{\it Boston University, Physics Department,\\
     590 Commonwealth Avenue,\\  Boston, MA 02215 USA}

\vspace{2cm}
{\bf ABSTRACT\ \ \ }
 \vspace{12pt}
\end{center}

\thispagestyle{empty}

\begin{minipage}{15.5cm}
In a wide class of extensions of the Standard Model there is a scalar
resonance with the quantum numbers of the usual Higgs boson but with
different couplings to fermions and gauge bosons. Using an effective
Lagrangian description, we examine the phenomenology of such a generic
\nh\ at the LHC.
In particular, we determine the circumstances under which such a
particle can be observed in its $ZZ$ decay mode and distinguished from
the Higgs boson of the Standard Model. We briefly comment on the energy
scale effectively probed at the LHC, if the nonstandard nature of an
observed Higgs particle can be asserted.

\end{minipage}
\end{titlepage}

\newpage
\renewcommand{\thefootnote}{\arabic{footnote}}
\setcounter{footnote}{0}
\setcounter{page}{2}
\baselineskip=18pt


\section{Introduction}
\hspace*{\parskip}  \label{sec-intro}
The operation of the next generation of high-energy colliders
(such as the LHC, LEP-II, NLC) within the coming decade is expected to
bring us closer to an understanding of the mechanism of electroweak
symmetry breaking. The minimal Standard Model (SM) is the simplest
possibility, but its confirmation requires the discovery of a neutral
scalar particle, the Higgs boson, with properties completely specified
once given its mass.
In the SM this is an undetermined parameter, and so far direct searches
have set a lower limit of about 60~GeV \cite{LEP}. An upper bound of
approximately 1~TeV has been suggested on the basis of ``triviality''
\cite{dashen}, and the validity of the perturbation expansion
\cite{kn:thacker}, which makes it likely that, if the SM Higgs boson
exists, it will be discovered at the next generation colliders.

However, it is widely believed \cite{dashen,thooft}
that the SM, despite its experimental success,
can not be complete and that new
physics, beyond the SM, should arise at some finite energy
scale $\Lambda$. If $\Lambda$ is very large, then the low-energy theory
would look like the SM, while if $\Lambda$ is low  (such as a few TeV),
then deviations should be expected and the properties of a Higgs-like
resonance (if present) could differ substantially from those predicted
in the context of the SM.
A resonance lighter than other massive degrees of freedom that shares
the quantum numbers of the SM Higgs boson but couples to the electroweak
gauge bosons and to fermions with nonstandard strength has been
generically called a ``Nonstandard Higgs'' boson \cite{vk1,vk2}.

Such an object is featured in a variety of models of electroweak
symmetry breaking; namely, some models with dynamical symmetry
breaking, such as Composite Higgs models
\cite{comphiggs,banks,dugan,kosower,vk2}
and ``top-condensate'' models \cite{topcon}, as well as linear models
with many fundamental scalars in which a mass gap exists between a light
scalar-isoscalar (under custodial isospin) particle and all other
resonances. If these models describe electroweak symmetry breaking, the
isoscalar resonance presumably will be the first to be discovered in a
collider experiment.
It is not clear {\it a priori}, however, whether its nonstandard
properties can be measured accurately enough to distinguish it from the
SM Higgs boson.

The question we wish to address in this paper is whether it will be
possible in future experiments at the Large Hadron Collider (LHC) to
detect a nonstandard Higgs boson and to differentiate it from the SM
Higgs.
As a model, we consider the most general low-energy effective Lagrangian
in terms of the usual $SU(2)_L\times SU(2)_R/SU(2)_V$\ symmetry breaking
pattern which, below the cutoff scale $\Lambda$, has the same spectrum as
the SM. The SM is a particular case and corresponds to the limit where
$\Lambda\rightarrow\infty$.
This Lagrangian is then used to explore the prospects of the
LHC to detect and distinguish a nonstandard from a Standard Higgs
boson.
In particular, for a variety of Higgs boson masses $m_H$ (assuming that
$m_H >2 m_Z$), we determine the values of couplings in the effective
Lagrangian for which this is possible by looking at the Higgs boson
decay mode $H\rightarrow ZZ \rightarrow l^+ l^- l^+ l^-$, where $l$ is
an electron or a muon.
It has been shown \cite{vk1,vk2} that if a scalar isoscalar resonance
is observed, then the measurement of its width offers the best way to
distinguish it from the SM Higgs.

The deviations from the SM couplings can be used within specific
models to estimate the scale $\Lambda$ of new physics, provided no
other nonstandard physics is discovered. As an indication, we have done
so for a number of simple Composite Higgs models.
This is similar in spirit to an early study by Kosower
\cite{kosower}, who also proposed the measurement of the width
as a tool to probe the compositeness scale within Composite Higgs
models. However, we performed a more detailed statistical analysis and
reached somewhat different (less optimistic) conclusions.

In the next section we review the theoretical framework and construct
the effective Lagrangian of the most general theory with a \nh. In
Section~3  we describe the calculation of the signal and $ZZ$
background cross-sections and discuss the issue of whether one can
discriminate between a \nh\ and its SM counterpart on the basis of a
width measurement. In particular, we derive the statistical significance
of a possible discrepancy between the result of such a measurement and
the SM expectation. Finally, Section~4 contains our conclusions.

\section{The Effective Lagrangian}

In this section, we briefly describe the construction of the most
general effective theory with a \nh\
\cite{vk1,vk2}.  The electroweak symmetry breaking sector at low
energies contains, besides the Goldstone bosons $w^a$ (which
become the longitudinal components of $W^\pm$ and $Z$), one extra scalar
particle $H$ (the \nh) 
with the quantum numbers of the SM Higgs boson\footnote{Throughout, we
shall call a generic \nh\ simply a Higgs boson unless otherwise
stated.}.

As in the SM, we assume that the Goldstone bosons arise from the
spontaneous breakdown  of a chiral \lr\ symmetry down to its diagonal
$SU(2)_V$ subgroup. As usual, $SU(2)_L$ is identified with the gauge
group $SU(2)_W$ and $SU(2)_R$ is the ``custodial'' symmetry whose
$\tau_3$ component is identified with hypercharge. The interactions
of the Goldstone bosons are described  conveniently by using a nonlinear
realization \cite{ccwz} of the chiral symmetry, in terms of the field
\be
\Sigma = \exp \left(\frac{i \vw \cdot \pauli}{v} \right) \ ,
    \label{eq-sigma}
\ee
where \pauli\ are the Pauli matrices corresponding to the broken $SU(2)$
generators, normalized so that $\trace \, (\tau^{a} \tau^{b}) = 2\,
\delta^{a b}$, and $v=246$~GeV is the weak scale.  Under the \lr\ chiral
symmetry the fields ($\Sigma, H$) transform as
\be
\Sigma \rightarrow L \Sigma R^\dagger \;\;\; , \;\;\;
H \rightarrow H  \label{eq-duo}
\ee
where $L$ and $R$ are $SU(2)_L$ and $SU(2)_R$ matrices respectively.
Then, the most general chirally invariant Lagrangian that describes the
interactions of the isoscalar $H$ with the Goldstone bosons $w^a$, to
lowest order in momentum, can be written as
\be
\cL = \frac{1}{4} \left(v^2 + 2 \xi v H + \xi^\prime H^2 +
 \ldots \right)\, \trace\,\left(\pdmu\Sigma^\dagger \PDmu\Sigma \right)
  \;\; + \;\;  \cL_H  \label{eq:efl}
\ee
where $\cL_H$ is the Lagrangian that describes the Higgs boson
self-interactions
\be
\cL_{H} = \frac{1}{2}(\pdmu H)^{2} - \frac{m_H^2}{2} H^2 -
 \frac{\lambda_3 v}{3 !} H^3  - \frac{\lambda_4}{4 !} H^4 - \ldots
  \label{eq:Lh}
\ee
and $\xi, \xi^\prime, \lambda_3 \; \mbox{\rm and}\; \lambda_4$ are
unknown coefficients. For simplicity, in eqs.~(\ref{eq:efl}) and
(\ref{eq:Lh}) we only show the leading terms,
with the ellipsis denoting higher powers in $H$.

The gauge bosons can be introduced by replacing the ordinary
derivative in \eq{eq:efl} by the covariant one, which, by virtue of the
transformation law (\ref{eq-duo}), takes the form
\be
D_{\mu}\Sigma = \partial_{\mu}\Sigma + i\frac{g}{2}\vec{\tau}\cdot
\vec{W_{\mu}}\Sigma - i\frac{g^\prime}{2} B_{\mu} \Sigma \tau_3
 \label{eq:cov}
\ee
where $g, g^\prime$ are the usual $SU(2)_W$ and $U(1)_Y$ gauge couplings
respectively. Hence the parameters $\xi, \xi^\prime$ etc,
represent the couplings of one or more nonstandard Higgs bosons
to a pair of weak gauge bosons $W^a_\mu$.

The fermions are incorporated into the effective Lagrangian as matter
fields \cite{ccwz}. We shall only consider the quarks of the third
family, since these will be the only important ones in our
phenomenological investigation. These fermions can be included in the
fields
\be
\psi_L=\left ( \begin{array}{c} t_L \\ b_L \end{array} \right )
\mbox{\hspace{0.5cm} ,\hspace{0.5cm}}
\psi_R=\left ( \begin{array}{c} t_R \\ b_R \end{array} \right )
   \label{eq-fermion}
\ee
which transform as $\psi_L \rightarrow L\psi_L$ and $\psi_R \rightarrow
R\psi_R$ under $SU(2)_L \times SU(2)_R$. Their interactions with the
scalars are given by
\be
\cL_{\Sigma f \bar{f}} = h_1(v +y_1 H + \ldots) \bar{\psi}_L\,\Sigma
   \,\psi_R +  h_2 (v +y_2 H + \ldots) \bar{\psi}_L \,\Sigma \, \tau_3
   \, \psi_R + {\rm h.c.}     \label{eq:lfer}
\ee
where  $h_1$ and $h_2$ correspond to Yukawa couplings and can be
replaced by the fermion masses, through $m_t = (h_1+h_2) v$ and
$m_b=(h_1-h_2)v$, while $y_1$ and $y_2$ are new unknown couplings.
Again, the ellipsis denotes higher powers in the Higgs field
which are not included in our analysis.
Using the explicit form (\ref{eq-fermion})
in eq.~(\ref{eq:lfer}), we can read off the Higgs boson couplings to the
top and bottom quarks:
\begin{eqnarray}
\cL_{Hf\bar{f}} & = &
(h_1 y_1 + h_2 y_2)\, H\bar{t}_L t_R + (h_1y_1- h_2 y_2)\, H \bar{b}_L b_R
\; + \;{\rm h.c.} \nonumber \\
 & \equiv & y_t \, (m_t/v)\, H\bar{t}_L t_R + y_b \, (m_b/v)\,
  H \bar{b}_L b_R \; +\; {\rm h.c.} \label{eq:htt}
\end{eqnarray}
Thus, this Lagrangian introduces two new unknown parameters $y_t,\, y_b$.

The SM is a particular case of the effective theory defined
above, with the only non-zero couplings being
\be
  \xi \, , \, \xi^\prime \; = \; 1 \;\;\; , \;\;\;
  \lambda_3\, , \, \lambda_4 \; = \; \frac{3 m_H^2}{v^2} \;\;\; , \;\;\;
  y_t \, , \, y_b \; = \; 1 \;\; .  \label{eq:smv}
\ee

For values of the couplings different from (\ref{eq:smv}) the effective
Lagrangian is nonrenormalizable and a cutoff is implicitly present.
We may estimate the order of magnitude of this cutoff
by determining the scale at which partial wave
unitarity is violated. The isospin-0 spin-0 partial wave amplitude, for
$s \gg m^2$, is \cite{vk1,cdg}
\be
a_{00}(s)=(1-\xi ^2) \frac{s}{16 \pi v^2}
\label{a00}
\ee
At a scale
\be
s=\frac{16\pi v^2}{|1-\xi ^2|}
\label{scut}
\ee
partial wave unitarity breaks down. Consequently the cutoff $\La$ of the
theory, physically associated with the scale at which new degrees of
freedom emerge, must lie at or below this scale.
 From eq.~(\ref{scut}) it becomes clear that the larger the deviation of
$\xi$ from its SM value of 1, the lower the energy scale at which new
physics is expected. If $\La \equiv 4\pi f$ is the scale of new physics,
then by inverting eq. (\ref{scut}), we can write, roughly,
\be
\xi ^2 = 1 + O\left (\frac{v^2}{f^2}\right )
\label{xieq1plus}
\ee
This is effectively the statement that nonrenormalizable couplings must
be suppressed by powers of the scale of new physics.
We thus expect similar relations to hold for the other couplings $y_t,
y_b, \xi^{\prime}$, etc.
Relations such as (\ref{xieq1plus}) are borne out by calculations in
specific models \cite{comphiggs,banks,dugan,kosower,vk2}.
Higher momentum contributions to the effective action can be
systematically taken into account using chiral perturbation theory
\cite{gl}, although, for simplicity, we do not include them in this
study.

It is clear that, since a \nh\ couples to the same channels as its SM
counterpart, its search strategy will be based on the same signatures.
Here, we shall assume that $m_H > 2
m_Z$. In this mass range, it turns out that  the nonstandard Higgs boson
properties are determined, to lowest order in perturbation theory, by
only two parameters, namely $\xi$ and $y_t$. The Higgs boson decay
width, for example, is given by
\begin{eqnarray}
\Gamma_H & = & \frac{m_H^3}{32\pi v^2} \xi^2 \, \left[ 2\,\sqrt{1-x_W}\,
(1-x_W+ \frac{3}{4} x_W^2) + \sqrt{1-x_Z}\, (1-x_Z+ \frac{3}{4} x_Z^2)
\right] \nonumber \\
 & & + \frac{3m_t^2 m_H}{8\pi v^2}
y_t^2 \left ( 1-\frac{4m_t^2}{m_H^2} \right )^{3/2}. \label{eq:fullwidth}
\end{eqnarray}
where $x_V=4m_V^2/m_H^2$, $V=W,Z$.
Here we are assuming
that the underlying short-distance dynamics acts so as not to
particularly enhance $y_b$ over $y_t$. Then, since $m_t \gg m_b$, only
the top quark couples significantly to the Higgs boson $H$, and we can
ignore the coupling to the bottom quark.
On the other hand, in the purely scalar sector, tree level amplitudes do
not depend on the parameters
$\xi^\prime, \lambda_3, \lambda_4$, etc.
The leading one-loop corrections to $W_L W_L$ scattering
and the Higgs boson decay width were computed in Ref.~\cite{vk1,vk2}
and, for phenomenological purposes, they can be incorporated in the
effective definition of $\xi$.
We now proceed to investigate the phenomenology of the model
presented above.

\section{Phenomenology}
\label{ch-chap4}

In this section, we explore the phenomenology of a \nh\ at the
LHC.  We shall only consider Higgs boson masses larger than the $ZZ$
threshold. In this mass range, the Higgs boson decays primarily to
gauge boson pairs and thus can be most effectively searched for in the
``gold-plated'' channel
\be
H \rightarrow ZZ \rightarrow l^+l^-l^+l^-   \label{gold}
\ee
where $l$ is an electron or a muon. This process has been discussed at
great length in the literature\footnote{Recent reviews appear, for
example, in Refs.~\cite{smphen,gem,rubbia}.}, and it is
expected that a Standard Higgs boson with mass
$m_H\leq 500$~GeV ($800$~GeV)
will be discovered at the LHC at an integrated luminosity of
10 fb$^{-1}$ (100 fb$^{-1}$).
The main question we try to answer in the analysis that follows is
whether a \nh\ of given mass $m_H$ and couplings $\xi$ and $y_t$ can be
detected at the LHC and distinguished from a SM Higgs boson of
the same mass. We present results for integrated luminosities of 10
fb$^{-1}$  and 100 fb$^{-1}$, while the center-of-mass energy
is assumed to be $\sqrt{s}=14.6$~TeV.
In our estimates of cross-sections we have used the ELHQ structure
functions, set 2 \cite{ehlq}.

The main background to the four lepton signal (\ref{gold}) comes
 from the Born process
\be
q\bar{q} \rightarrow ZZ \rightarrow l^+l^-l^+l^-
\label{qqzz}
\ee
We calculated this background imposing the following cuts on the
rapidity and transverse momentum of the $Z$ bosons
\be
|\eta^Z|<2 \mbox{\hspace{1.5cm}};\mbox{\hspace{1.5cm}} p_T^Z >
\frac{1}{4} \sqrt{m_{ZZ}^2-4\,m_Z^2} \;\;\; .   \label{zzcuts}
\ee
The rapidity cut for the $Z$'s translates, approximately,
into a cut of 2.5
for the rapidities of the final state leptons. We included part
of the QCD corrections to this process, through a ``K-factor''
\cite{kqq}
\be
K=1+\frac{8}{9}\pi \alpha _s(m_H) .
\ee

An irreducible $ZZ$ background also arises from gluon fusion through a
quark loop (the `box' diagram)
\be
gg \rightarrow ZZ \;\; .
\ee
In fact, this process interferes with the resonant Higgs boson exchange process
\be
gg \rightarrow H \rightarrow ZZ
\label{gghzz}
\ee
where the Higgs boson is produced through a top quark loop. An exact
calculation for the SM \cite{glover} has shown that the
effect of the interference is rather small, for most of the range of
masses we consider. Towards the upper end of this range, however, (that
is $m_H \approx 800$~GeV) the increase in the cross-section caused by the
interference term may become sizeable (it is constructive interference).
We ignored this effect, and thus our estimates of the signal rate are
somewhat conservative for large masses. We did take into account,
however, the contribution of the `box' diagram to the background, which
amounts to approximately 50\% of the Born process (\ref{qqzz}), by
scaling the cross-section of the latter by 1.5.

There are also reducible four-lepton backgrounds, primarily from
$t\bar{t}$ production. It has been argued \cite{smphen,nisati}
that, with appropriate isolation cuts and the expected $Z$ mass
resolution capability at the LHC, these backgrounds can be reduced to
well below the irreducible background levels. We shall therefore ignore
them in this study. However, we have taken into account a 10\% loss of
signal rate due to these cuts \cite{nisati}.

The main mechanism for $Z$-boson pair production through a Higgs boson
is the process (\ref{gghzz}). The rate for this process depends on
the top quark mass, which is chosen here to be
$m_t=170$~GeV \cite{CDF}, and also on the nonstandard Yukawa
coupling $y_t$. In the Standard Model, for such a top-quark
mass, the gluon fusion is the dominant production mechanism for all
Higgs boson masses up to 1 TeV.  Leading QCD corrections to this process
have been included by multiplying the cross section by another ``K-factor''
\cite{djouadigl,gr,kz}
\be
K = 1+\left (\frac{11}{2}+\pi^2 \right ) \frac{\alpha_s(m_H)}{\pi}
\;\; .
\ee

A second production mechanism for $Z$ pairs through a Higgs boson is
gauge boson fusion
\be
qq \rightarrow qqH \rightarrow  qqZZ  \;\; .  \label{wfusion}
\ee
We computed the cross-section for this process by using the
effective-$W$ approximation \cite{effw,chanowitz}.
The scattering amplitudes are calculated at tree-level in
the gauge theory from the Lagrangian of Section~2.
The cross-section is obtained by
folding the amplitudes with the luminosities of the $W$'s and $Z$'s
inside a quark. Both transverse and longitudinal polarizations are
included using the distribution functions of Ref.~\cite{repko} (see
also \cite{baur}). (The subleading terms in the expressions for these
functions depend on the characteristic energy scale of the process under
consideration, taken here to be $Q^2=m^2_{ZZ}/4$.)
The contribution from $W_L W_L (Z_L Z_L)$ fusion, which is the least
affected by the choice of $Q^2$, is dominant for energies around
the peak, since this amplitude is most sensitive to the existence of the
Higgs resonance, while the $W_L W_T +W_T W_T $ fusion prevails
outside this region. The contribution to the cross-section
 from the interaction of the gauge bosons that does not
involve the exchange of the Higgs resonance should in fact be
considered as a background \cite{rogerio}. We have calculated this
background in the effective theory with $\xi=0$, and
subtracted it from the cross-section of the process (\ref{wfusion}) in
our estimates for the signal.
We should also remark that in the calculation of both processes
(\ref{wfusion}) and (\ref{gghzz}), the $s$-channel Higgs boson exchange
diagram is unitarized by including the ``running'' Higgs boson decay
width\footnote{The running width is obtained through the relation
${\rm Im}\, \Pi_H(s)=-\sqrt{s}\,\Gamma_H(m_H^2=s)$.}
in the propagator. This prescription (which can be justified only in the
resonance region) differs from other ones, such as including a
constant width, by effects which are formally of higher order in
$\lambda \equiv m_H^2/2 v^2$. However, for $WW$ scattering, it was shown
\cite{kn:valencia92} that it is better to use an energy dependent
width, because partial wave amplitudes stay closer to the unitarity
circle. In terms of event rates, we found that, for gluon fusion, the
two prescriptions differ by at most 10$\%$ for a heavy and wide
resonance (see also Ref.~\cite{kn:moriond}).

In Tables~\ref{t1}--\ref{t6} we present our results for the event rates
and the statistical significance of Higgs boson signals for various
values of $\xi$ and $y_t$. The resonance will have an effective width
$\Gamma_{eff}$ determined by the physical Higgs boson width and by the
mass resolution of the detector:
\be
\Gamma_{eff}=\sqrt{(\Delta m_H)^2+\Gamma_H^2} \;\;\; .
\label{effwidth}
\ee
We assume that the Higgs boson mass can be resolved to within $\Delta m_H
= 4\%\, m_H$ \cite{smphen,nisati}. We have also assumed an
identification efficiency of $90 \%$ per lepton \cite{kn:seez}.
The number of events is measured in a mass bin of width $\Gamma _{eff}$
centered at the resonance peak. We note that, in all cases examined,
this resonance region lies reasonably below the
cutoff where the effects of non-unitarity become appreciable.
The statistical significance of the signal is determined by computing
the Poisson probability that the signal is due to background
fluctuations \cite{gem}.
We also note that for large masses and widths the resonance peak occurs,
in general, at a lower energy than the nominal Higgs boson mass $m_H$,
due to the interference with the non-resonant terms, the energy
dependence of the (running) width and the effect of the falling
distribution functions. For example, if $m_H=800$ (600)~GeV
and $\xi=y_t=1$, the
maximum of the signal cross-section occurs at approximately 730 (585)~GeV.
In our results, the Higgs boson mass quoted refers
to $m_H$ rather than the resonance mass.
The results presented for the event rates at the high luminosity
(100 fb$^{-1}$) were obtained from those at low luminosity
(10 fb$^{-1}$) by scaling by a factor of 10. This is not, strictly
speaking, a correct procedure, because of the problems a higher
luminosity environment may pose (such as deterioration in the energy
resolution) \cite{gem,kn:seez}. A full detector simulation is needed
in order to assess the magnitude of these effects.
Consequently, our results for a luminosity of 100~fb$^{-1}$ should be
regarded as rather optimistic.

 From these tables it can be seen that a nonstandard Higgs resonance
may be distinguished in principle from the SM Higgs boson by a
comparison of its width and total cross-section to the Standard Model
predictions. Before we decide whether this can be achieved in practice,
we need to know the expected accuracy of a width measurement,
as well as the theoretical uncertainties in
the calculation of the width and the cross-section.
There are few theoretical uncertainties in the calculation
of the SM Higgs boson width. Higher order corrections to both gauge boson and
fermion decay modes have been computed \cite{kn:wcorr,kn:kks}
and have been found to increase the full width by approximately $15 \%$.
We chose here not to include this correction, but this does not alter
our conclusions. (It will simply change the effective SM value
of $\xi$ and $y_t$ to a value slightly different from 1.) For the
purposes of deciding whether an observed resonance is consistent with
the Standard Model predictions, what matters is to know the latter
precisely enough, which we do. Similarly, we have chosen not
to include radiative corrections to the width of a \nh\ since these can
be incorporated into the definition for $\xi$ and $y_t$
\cite{vk2}.
In contrast, the accuracy of the cross-section calculation is
compromised by the imprecise knowledge of structure functions (amounting
to perhaps 30\% for Higgs boson production \cite{gem}),
our various approximations (such as the effective-$W$ scheme or the
neglect of the interference effects of the `box' diagram in $ZZ$
production) as well as further corrections beyond the included QCD effects.
Consequently, if a Higgs-like resonance is discovered, a comparison of
its width to the Standard Model prediction offers the best way to probe
its nature.

The systematic uncertainty in the measurement of the width arising from
smearing may be corrected for by using \eq{effwidth}. This will
be an accurate procedure only if $\Gamma_H \simgt \Delta m_H$.
The statistical error involved in the measurement of the width warrants
a more detailed discussion:
Suppose that a Higgs resonance is observed at a mass $m_H$
and its width measured and found to differ from the expected \sm value
$\Gamma_{SM}$. We wish to attach a statistical significance to this
deviation. This statistical significance can be derived from the
probability density function according to which the possible
measurements of the Standard
Higgs boson width are distributed. (Any measured quantity is a statistical
variable and, as such, obeys some probability distribution function.)
To obtain the probability distribution we performed a large number of
numerical experiments simulating the possible outcomes of an actual
experiment. The procedure adopted was the following: the $ZZ$
invariant-mass
range of interest was divided in 4-GeV bins. In each of them the
total number of events was generated according
to a Poisson distribution with mean $N_S + N_B$, where $N_S$, $N_B$ are
the SM signal and background events respectively, expected in that bin.
Assuming that the continuum background is known (e.g. from independent
experiments) we subtracted the expected background $N_B$ in each bin.
The resulting distribution represents the signal with an additional
noise due to background fluctuations. The mass and the width were
obtained by fitting this data with a function of the form
\be
\frac{m^4}{E}\; \frac{e^{-E/E_0}}{(E^2-m^2)^2+m^2 \Gamma ^2}
\label{fit}
\ee
where $E$ is the invariant mass of the $Z$ pair and $m,\Gamma$ are the
parameters of the fit. The exponential encodes the effect of the falling
parton distribution functions, while in the expression for the
cross-section, factors other than the propagator have a rough $m^4/E$
dependence. The value of the constant $E_0$ was fixed from the exact
(lowest-order) cross-section for the process
\be
pp(gg) \rightarrow H \rightarrow ZZ
\ee
The best fit occurs for $E_0=283.8\, \gev $.

Repeating this experiment a large number of times, we were able to
obtain the probability density, the mean $<\Gamma_{eff}>$
and the standard deviation $\delta \Gamma_{eff}$. As mentioned earlier,
the physical Higgs boson width can be recovered from the measured, or
``effective'', width $\Gamma_{eff}$ through \eq{effwidth}. In
particular, the spread $\delta \Gamma_{SM}$ that corresponds to one
standard deviation $\delta \Gamma_{eff}$ is given by
\be
\delta \Gamma_{SM} = \delta \Gamma_{eff}
\frac{1}{\sqrt{1-(\Delta m_H/\Gamma_{eff} )^2}}
\label{onesigma}
\ee
Thus, if a resonance of (physical) width $\Gamma_H \neq \Gamma_{SM}$ is
observed, the statistical significance $S$ associated with this
discrepancy is given by the number of standard deviations that
$\Gamma_H$ lies away from $\Gamma_{SM}$:
\be
S=\frac{|\Gamma_H-\Gamma_{SM}|}{\delta \Gamma_{SM}}
\ee
Because the underlying statistics is Poisson distributed, we
expect that the standard deviation $\delta \Gamma_{eff}$ scales with the
total number of events $N$ like
\be
\frac{\delta \Gamma_{eff}}{\Gamma_{eff}}=\frac{c}{\sqrt{N}}
\ee
In the limit of large $N$ and negligible background, $c$ is a
constant\footnote{In the narrow width approximation, that is, ignoring
the exponential factor $\exp(-E/E_0)$ in eq. (\ref{fit}), we found $c=1.25$.
Note that this result is process-independent and reflects only
the underlying statistics. It could therefore be applied to other
analyses as well.}. In general, though, $c$ is a function of both the
signal $N$ and the background $B$
(and, as can be expected, increases with increasing $B$ or
decreasing $N$). For poor statistics and wide objects (for instance in
the case $m_H=800\, \gev$), the width can hardly be measured, even if a
statistically significant signal can be obtained.

In Tables~\ref{t7}--\ref{t8} we display, for various masses and three
representative values of $y_t$, namely $y_t^2=0.5,1$ and 2, the range of
values of $\xi$ for which the nonstandard Higgs boson is observable and
distinguishable from the SM Higgs. Results are
presented for integrated luminosities of 10~fb$^{-1}$ and 100~fb$^{-1}$.
The criteria used in compiling these tables are the following: For a
signal to be declared ``observable'' we require that it consists of at
least 10 events and that its statistical significance is greater than
5$\sigma$. For it to be distinguishable from the SM Higgs boson, we
require that its width $\Gamma_H$ differ from the Standard Model value
by at least three standard deviations as defined by \eq{onesigma}. If
this criterion is not satisfied, one could in principle examine the
signal event rate. However, given the large uncertainty in the
theoretical calculation, we opted not to use this information.

\section {Conclusions}

Our conclusions are consistent with the expectation that a SM Higgs boson will
be detected at the LHC in this channel
provided its mass is less than about 500~GeV (at
10~fb$^{-1}$) or 800~GeV (at 100~fb$^{-1}$). As $y_t$ becomes smaller or
larger than unity, this mass range will shrink or expand.
For example, at $y_t^2=0.5$ and
$\xi=1$ the respective mass ranges  at the low and
high luminosity options considered are 330~GeV $\simlt m_H \simlt $
430~GeV and $2M_Z \simlt m_H \simlt $ 680~GeV respectively.
We observe further that at
10~fb$^{-1}$, only models with relatively large $\xi$ can
be differentiated from the Standard Model. This is primarily due to the
low statistics and the consequent imprecision in the width measurement.
It might be possible, however, to improve the statistics by a less
strict set of cuts on the final state leptons (or $Z$'s). The situation
is considerably better at 100~fb$^{-1}$, as can be seen from
Table~\ref{t8}.

In certain cases where $\xi$ is small, the nonstandard Higgs boson is
too narrow to be resolved, even though a SM Higgs of the same mass is
not. In this case one could tell that the Higgs boson is nonstandard by
comparing the detector resolution to the expected SM width, but it is not
possible to determine a value for $\xi$.

As we emphasized earlier, the deviation of the values of the parameters
$\xi$ and $y_t$ from unity is a measure of the cutoff $\La$,
which can be thought of as an upper bound to the scale of new
physics. Precise relations, however, are model-dependent. In the context
of specific models, the results presented in Tables 7-8 reveal the
energy scale that the LHC will be able to probe.
For example, if $m_H=500$~GeV, where the sensitivity of the LHC to the
measurement of $\xi$ is about 30\% (see Table~\ref{t8}), the scale
probed is $\La=4.3$~TeV in the $SU(3)_L\times SU(3)_R/SU(3)_V$ model of
Ref.~\cite{banks} where $\xi^2=1- v^2/f^2$,
$\La=2.2$~TeV in the $SU(5)/SO(5)$ model \cite{dugan} where $\xi^2
=1-(v^2/4f^2)$, and finally $\Lambda=16$~TeV in the $SU(4)/SU(2)\times
SU(2)$ model \cite{kosower} in which $\xi^2 =1-(4v^2/f^2)$. In the
above, $\La=4\pi f$ is the compositeness scale of the underlying new
strong dynamics and $v=246$~GeV, while we have assumed $y_t=1$ in
all of these cases. In a general Two-Higgs-Doublet model where a gap
exists between the mass $m_H$ of the lightest neutral state and that of
the heavier (nearly degenerate) scalars ($M$, say), the parameter $\xi$
generally approaches its SM value faster: $\xi^2=1-O(m_H^4/M^4)$; our
results indicate that, in this case, it will be very
hard to determine the existence of a non-minimal scalar
sector solely from a measurement of
the width of the observed resonance (see also Ref.~\cite{haber}).

\vspace{0.9cm}
\noindent
{\Large\bf Acknowledgements}

We thank R.~S.~Chivukula, M.~Golden, K.~Lane and B.~Zhou for useful
conversations. This work was supported in part under NSF contract
PHY-9057173 and DOE contract DE-FG02-91ER40676.

\clearpage

\newpage
\begin{table}
\begin{center}
\begin{tabular}{|c|l|c|l|c|l|c|}     \hline
 & \multicolumn{2}{c|}{$m_H=200$~GeV} &
\multicolumn{2}{c|}{$m_H=350$~GeV}  & \multicolumn{2}{c|}{$m_H=500$~GeV}
\\ \hline
$\xi$  & Ev. (Sign.)& Width & Ev. (Sign.)& Width &
Ev. (Sign.)& Width  \\ \hline
0.25 & 39 (5.6) & 0.09 & 26 (7.6) & 1.21 & 3.7 & 14.5 \\
\hline
0.50 & 40 (5.8) & 0.36 & 32 (9.0) & 4.02 & 11 (4.8) & 24.2 \\
\hline
0.75 & 42 (6.0) & 0.80 & 34 (9.0) & 8.71 & 14 (5.2) & 40.5 \\
\hline
1.00 & 45 (6.4) & 1.41 & 36 (8.8) & 15.3 & 16 (5.2) & 63.3  \\
\hline
1.25 & 49 (6.9) & 2.21 & 43 (9.3) & 23.7 & 18 (5.0) & 92.5  \\
\hline
1.50 & 54 (7.4) & 3.18 & 43 (8.5) & 34.0 & 20 (4.6) & 128  \\
\hline
1.75 & 59 (8.2) & 4.32 & 45 (8.2) & 46.2 & 22 (4.4) & 170  \\
\hline
2.00 & 66 (8.9) & 5.65 & 46 (7.6) & 60.3 & 24 (4.0) & 219  \\
\hline
\end{tabular}
\caption{Event rates and decay widths for various Higgs boson masses
$m_H$ and $\xi$ at the LHC at a luminosity of 10~fb$^{-1}$ for standard
top Yukawa coupling and $m_t=170$~GeV. The statistical significance
is also shown for signals consisting of more than 10
events.}
\label{t1}
\end{center}
\end{table}


\begin{table}
\begin{center}
\begin{tabular}{|c|l|c|l|c|l|c|l|c|}     \hline
 & \multicolumn{2}{c|}{$m_H=200$~GeV} &
\multicolumn{2}{c|}{$m_H=400$~GeV}  & \multicolumn{2}{c|}{$m_H=600$~GeV}
 & \multicolumn{2}{c|}{$m_H=800$~GeV}\\ \hline
$\xi$  & Ev. (Sg.)& $\Gamma_H$ & Ev. (Sg.)& $\Gamma_H$&
Ev. (Sg.)& $\Gamma_H$ & Ev. (Sg.)& $\Gamma_H$\\ \hline
0.25 & 389 (17.9) & 0.09 &  98 (12.7) & 4.85 & 22 (5.5) & 25.1 & 6.9 &
48.6 \\
\hline
0.50 & 403 (18.5) & 0.36 & 205 (22.9) & 9.40 & 50 (9.3) & 42.8 & 14
(4.0) & 92.9 \\
\hline
0.75 & 418 (19.0) & 0.80 & 264 (25.1) & 17.0 & 68 (10.2) & 72.4 & 21
(4.2) & 167 \\
\hline
1.00 & 446 (20.2) & 1.41 & 340 (28.1) & 27.6 & 82 (10.1) & 114 & 28
(4.0) &270 \\
\hline
1.25 & 486 (21.8) & 2.21 & 352 (26.5) & 41.2 & 96 (9.5) & 167 & 36 (3.5)
& 403 \\
\hline
1.50 & 535 (23.7) & 3.18 & 368 (24.6) & 57.9 & 111 (8.6) & 232 & 44
(2.8) & 565 \\
\hline
1.75 & 593 (25.9) & 4.32 & 387 (22.8) & 77.6 & 126  (8.0) & 309 &
\multicolumn{2}{c|}{*} \\
\hline
2.00 & 659 (28.4) & 5.65 & 406 (21.8) & 100  & 140 (6.4) & 398 &
\multicolumn{2} {c|}{*} \\
\hline
\end{tabular}
\caption{Event rates and decay widths for various Higgs boson masses $m_H$ and
$\xi$ at the LHC at a luminosity of 100~fb$^{-1}$ for standard top Yukawa
coupling and $m_t=170$~GeV. The statistical significance
is also shown for signals consisting of more than 10
events. The star indicates that the Higgs boson is too wide to be considered a
resonance ($\Gamma_H \sim m_H$).}
\label{t2}
\end{center}
\end{table}

\clearpage


\begin{table}
\begin{center}
\begin{tabular}{|c|l|c|l|c|l|c|}     \hline
 & \multicolumn{2}{c|}{$m_H=200$~GeV} &
\multicolumn{2}{c|}{$m_H=350$~GeV}  & \multicolumn{2}{c|}{$m_H=500$~GeV}
\\ \hline
$\xi$  & Ev. (Sign.)& Width & Ev. (Sign.)& Width &
Ev. (Sign.)& Width  \\ \hline
0.25 & 20 (3.0) & 0.09 & 15 (4.9) & 1.07 & 3.1 & 8.87 \\
\hline
0.50 & 20 (3.1) & 0.36 & 17 (5.4) & 3.89 & 6.1 & 18.6 \\
\hline
0.75 & 22 (3.3) & 0.80 & 18 (5.4) & 8.57 & 8.8 & 34.9 \\
\hline
1.00 & 25 (3.7) & 1.41 & 20 (5.5) & 15.1 & 10 (3.5) & 57.6  \\
\hline
1.25 & 28 (4.2) & 2.21 & 25 (6.0) & 23.6 & 11 (3.4) & 86.9  \\
\hline
1.50 & 33 (4.9) & 3.18 & 26 (5.7) & 33.9 & 13 (3.2) & 123  \\
\hline
1.75 & 39 (5.7) & 4.32 & 28 (5.6) & 46.1 & 14 (3.1) & 165  \\
\hline
2.00 & 46 (6.5) & 5.65 & 31 (5.3) & 60.1 & 16 (2.8) & 214  \\
\hline\hline
SM   & 45 (6.4) & 1.41 & 36 (8.8) & 15.3 & 16 (5.2) & 63.3 \\ \hline
\end{tabular}
\caption{Event rates and decay widths for various masses and $\xi$ at
the LHC at a luminosity of 10~fb$^{-1}$ for nonstandard top Yukawa
coupling $y_t^2=0.5$ and $m_t=170$~GeV. The statistical
significance is also shown for signals consisting of more
than 10 events. }
\label{t3}
\end{center}
\end{table}


\begin{table}
\begin{center}
\begin{tabular}{|c|l|c|l|c|l|c|l|c|}     \hline
 & \multicolumn{2}{c|}{$m_H=200$~GeV} &
\multicolumn{2}{c|}{$m_H=400$~GeV}  & \multicolumn{2}{c|}{$m_H=600$~GeV}
 & \multicolumn{2}{c|}{$m_H=800$~GeV}\\ \hline
$\xi$  & Ev. (Sg.)& $\Gamma_H$ & Ev. (Sg.)& $\Gamma_H$ &
Ev. (Sg.)& $\Gamma_H$  & Ev. (Sg.)& $\Gamma_H$ \\ \hline
0.25 & 198 (9.6) & 0.09 & 75 (10.0) & 3.18 & 15 (4.1) & 15.5 & 5.9 &
31.7  \\
\hline
0.50 & 204 (9.9) & 0.36 & 126 (15.5) & 7.73 & 35 (7.4) & 33.2 & 10 (3.0)
& 76.0 \\
\hline
0.75 & 218 (10.5) & 0.80 & 152 (17.0) & 15.3 & 43 (7.2) & 62.8 & 13
(3.0) & 150\\
\hline
1.00 & 245 (11.7) & 1.41 & 190 (18.3) & 25.9 & 50 (7.0) & 104  & 18
(2.9) & 253\\
\hline
1.25 & 285 (13.5) & 2.21 & 202 (17.3) & 39.5 & 62 (6.6) & 158  & 25
(2.5) & 386\\
\hline
1.50 & 334 (15.6) & 3.18 & 220 (16.6) & 56.2 & 74 (6.5) & 223  & 32
(2.2) & 549\\
\hline
1.75 & 392 (18.0) & 4.32 & 240 (15.5) & 75.9 & 87 (5.7) & 299  &
\multicolumn{2}
{c|}{*} \\
\hline
2.00 & 458 (20.7) & 5.65 & 268 (15.2) & 98.6 & 100 (4.7) & 388  &
\multicolumn{2}
{c|}{*} \\
\hline\hline
SM   & 446 (20.2) & 1.41 & 340 (28.1) & 27.6 & 82 (10.1)& 114  & 28
(4.0) & 270 \\
\hline
\end{tabular}
\caption{Event rates and decay widths for various masses and $\xi$ at
the LHC at a luminosity of 100~fb$^{-1}$ for nonstandard  Higgs-top
Yukawa coupling $y_t^2=0.5$ and $m_t=170$~GeV. The statistical
significance is also shown for signals consisting of more
than 10 events. The star indicates that the Higgs boson is too wide to be
considered a resonance ($\Gamma_H \sim m_H$).}
\label{t4}
\end{center}
\end{table}

\clearpage


\begin{table}
\begin{center}
\begin{tabular}{|c|l|c|l|c|l|c|}     \hline
 & \multicolumn{2}{c|}{$m_H=200$~GeV} &
\multicolumn{2}{c|}{$m_H=350$~GeV}  & \multicolumn{2}{c|}{$m_H=500$~GeV}
\\ \hline
$\xi$  & Ev. (Sign.)& Width & Ev. (Sign.)& Width &
Ev. (Sign.)& Width  \\ \hline
0.25 & 75 (9.9) & 0.09 & 42 (11.1) & 1.48 & 4.9 & 25.7 \\
\hline
0.50 & 79 (10.5) & 0.36 & 59 (14.3) & 4.29 & 14 (5.3) & 35.5 \\
\hline
0.75 & 82 (10.7) & 0.80 & 64 (14.5) & 8.98 & 21 (6.8) & 51.7 \\
\hline
1.00 & 85 (11.0) & 1.41 & 68 (14.4) & 15.5 & 26 (7.1) & 74.5  \\
\hline
1.25 & 89 (11.5) & 2.21 & 78 (14.8) & 24.0 & 29 (7.2) & 104   \\
\hline
1.50 & 94 (12.0) & 3.18 & 75 (13.5) & 34.3 & 33 (6.9) & 140   \\
\hline
1.75 & 99 (12.6) & 4.32 & 77 (12.7) & 46.5 & 35 (6.5) & 182   \\
\hline
2.00 & 106 (13.3) & 5.65 & 76 (11.6) & 60.5 & 38 (5.7) & 231   \\
\hline\hline
SM   & 45 (6.4) & 1.41 & 36 (8.8) & 15.3 & 16 (5.2) & 63.3 \\ \hline
\end{tabular}
\caption{Event rates and decay widths for various masses and $\xi$ at
the LHC at a luminosity of 10~fb$^{-1}$ for nonstandard Higgs-top
Yukawa coupling $y_t^2=2$ and $m_t=170$~GeV. The statistical
significance is also shown for signals consisting of more
than 10 events. }
\label{t5}
\end{center}
\end{table}


\begin{table}
\begin{center}
\begin{tabular}{|c|l|c|l|c|l|c|l|c|}     \hline
 & \multicolumn{2}{c|}{$m_H=200$~GeV} &
\multicolumn{2}{c|}{$m_H=400$~GeV}  & \multicolumn{2}{c|}{$m_H=600$~GeV}
 & \multicolumn{2}{c|}{$m_H=800$~GeV}\\ \hline
$\xi$  & Ev. (Sg.)& $\Gamma_H$ & Ev. (Sg.)& $\Gamma_H$ &
Ev. (Sg.)& $\Gamma_H$  & Ev. (Sg.)& $\Gamma_H$ \\ \hline
0.25 & 751 (31.7) & 0.09 & 116 (14.6) & 8.19 & 23 (4.8) & 44.2 & 7.6 &
82.4  \\
\hline
0.50 & 795 (33.3) & 0.36 & 301 (28.9) & 12.7 & 65 (10.4) & 61.9 & 20
(4.8) & 127 \\
\hline
0.75 & 816 (34.0) & 0.80 & 497 (40.2) & 20.3 & 100 (13.2) & 91.5 & 32
(5.7) & 201 \\
\hline
1.00 & 846 (35.0) & 1.41 & 546 (40.3) & 30.9 & 131 (14.0) & 133 & 44
(5.7) & 304 \\
\hline
1.25 & 886 (36.4) & 2.21 & 611 (39.7) & 44.5 & 156 (13.8) & 186  & 56
(5.0) & 437  \\
\hline
1.50 & 936 (38.1) & 3.18 & 636 (37.4) & 61.2 & 178 (12.7) & 251  & 67
(3.9) & 599 \\
\hline
1.75 & 995 (40.1) & 4.32 & 643 (34.7) & 80.1 & 198 (11.1) & 328  &
\multicolumn{2}{c|}{*} \\
\hline
2.00 & 1060 (42.3) & 5.65 & 675 (33.0) & 104  & 214 (9.0) & 417  &
\multicolumn{2}{c|}{*} \\
\hline\hline
SM   & 446 (20.2) & 1.41 & 340 (28.1) & 27.6 & 82 (10.1) & 114  & 28
(4.0) & 270  \\
\hline
\end{tabular}
\caption{Event rates and decay widths for various masses and $\xi$ at
the LHC at a luminosity of 100~fb$^{-1}$ for nonstandard  Higgs-top
Yukawa coupling $y_t^2=2$ and $m_t=170$~GeV. The statistical
significance is also shown for signals consisting of more
than 10 events. The star indicates that the Higgs boson is too wide to be
considered a resonance ($\Gamma_H \sim m_H$).}
\label{t6}such an object
\end{center}
\end{table}

\clearpage

%
\begin{table}
\begin{center}
\begin{tabular}{|c|c|c|c|}     \hline
$m_H$~(GeV) & \multicolumn{3}{c|}{Range of $\xi$} \\ \cline{2-4}
            & $y_t^2=0.5$ & $y_t^2=1$ & $y_t^2=2$ \\ \hline
300 & -- & -- & -- \\ \hline
350 & $\xi \simgt 1.70$ & $\xi \simgt 1.70$ & $\xi \simgt 1.70$ \\
\hline
400 & $1.60 \simlt \xi \simlt 1.75$ & $\xi \simgt 1.60$ &
                $\xi \simgt 1.55$  \\ \hline
450 & -- & $\xi \simgt 1.70$  & $\xi \simgt 1.65$ \\ \hline
500 & -- & -- & $\xi \simgt 1.80$ \\ \hline
\end{tabular}
\caption{Range of values of the parameter $\xi$ for which the
nonstandard Higgs boson resonance is both observable and distinguishable
 from the Standard Higgs at 10~fb$^{-1}$ of luminosity and three values
of the nonstandard Yukawa coupling $y_t$.
The range $0 < \xi \leq 2.0$ has been explored.}
\label{t7}
\end{center}
\end{table}

%
\begin{table}
\begin{center}
\begin{tabular}{|c|c|c|c|}     \hline
$m_H$~(GeV) & \multicolumn{3}{c|}{Range of $\xi$} \\ \cline{2-4}
            & $y_t^2=0.5$ & $y_t^2=1$ & $y_t^2=2$ \\ \hline
200 & --  & -- & -- \\ \hline
300 & $\xi \simgt 1.40$ & $\xi\simgt 1.40$ & $\xi \simgt 1.40$ \\ \hline
400 & $0.20\simlt \xi\simlt 0.45$ & $0.20\simlt \xi\simlt 0.35$ &
    $\xi\simgt 1.15$ \\
    & $\xi \simgt 1.25$ & $\xi\simgt 1.20$ & \\ \hline
500 & $0.25\simlt \xi\simlt 0.70$ & $0.20\simlt \xi\simlt 0.60$ &
    $0.20\simlt \xi\simlt 0.40$ \\
    & $\xi \simgt 1.35$ & $\xi\simgt 1.30$ & $\xi\simgt 1.20$ \\ \hline
600 & $\xi \simgt 1.45$ & $\xi \simgt 1.40$ &
    $ \xi\simgt 1.35$ \\ \hline
700 & -- &  $1.60 \simlt \xi\simlt 1.90$ & $\xi\simgt 1.55$
            \\ \hline
800 & -- & -- & -- \\  \hline
\end{tabular}
\caption{Range of values of the parameter $\xi$ for which the
nonstandard Higgs boson resonance is both observable and distinguishable
 from the Standard Higgs at 100~fb$^{-1}$ of luminosity and three values of
the nonstandard Yukawa coupling $y_t$.
The range $0 < \xi \leq 2.0$ has been explored.}
\label{t8}
\end{center}
\end{table}


\clearpage

\newpage
\begin{thebibliography}{99}

\bibitem{LEP} ALEPH Collaboration, D. Decamp et al., \pr{\bf 216}
 (1992) 253;\\
 DELPHI Collaboration, P. Abreu et al., \np{\bf 373} (1992) 3;\\
 L3 Collaboration, O. Adriani et al., \pl{\bf 303} (1993) 391;\\
 OPAL Collaboration, M. Akrawy et al., \pl{\bf 253} (1991) 511.
\bibitem{dashen} R.~Dashen and H.~Neuberger, \prl {\bf 50} (1983) 1897.
\bibitem{kn:thacker}B.W. Lee, C. Quigg and H. Thacker, \prd{\bf 16}
 (1977) 1519.
\bibitem{thooft} G.'t Hooft, in: {\em Recent Developments in Gauge
 Theories}, edited by G.~'t~Hooft (Plenum Press, New York, 1980).
\bibitem{vk1}R.~S.~Chivukula and V.~Koulovassilopoulos, \pl{\bf 309}
 (1993) 371.
\bibitem{vk2} V.~Koulovassilopoulos and R.~S.~Chivukula, Phys. Rev.
 {\bf D50} (1994) 3218.

\bibitem{comphiggs}D. B. Kaplan and H. Georgi, \pl{\bf 136} (1984) 183;\\
  D. B. Kaplan, S. Dimopoulos and H. Georgi, \pl{\bf 136} (1984) 187.
\bibitem{banks} T. Banks, \np{\bf 243} (1984) 125.
\bibitem{dugan}
   H. Georgi and D. B. Kaplan, \pl{\bf 145} (1984) 216;\\
  M. J. Dugan, H. Georgi and D. B. Kaplan, \np{\bf 254} (1985) 299.
\bibitem{kosower} D. A. Kosower, in {\em Physics of the Superconducting
 Supercollider}, Proceedings of the 1986 Summer Study on the Physics of
 the SSC, Snowmass 1986, edited by R.~Donaldson and J.~Marx.
\bibitem{topcon} Y.~Nambu, Enrico Fermi Institute Preprint EFI 88-39;\\
  V.~A.~Miransky, M.~Tanabashi, and K.~Yamawaki, \pl{\bf 221} (1989)
  177;  Mod. Phys. Lett. {\bf A4} (1989) 1043;\\
  W.~A.~Bardeen, C.~T.~Hill, and M.~Lindner, \prd{\bf 41} (1990) 1647.

\bibitem{ccwz}S. Coleman, J. Wess and B. Zumino, \pr {\bf 177}
   (1969) 2239;\\ C.~G.~Callan, S.~Coleman, J.~Wess and B.~Zumino,
               \pr {\bf 177} (1969) 2247.
\bibitem{cdg}R. S. Chivukula, M. Dugan and M. Golden, \pl{\bf 336}
      (1994) 62.
\bibitem{gl}S. Weinberg, Physica 96 A (1979) 327;\\
J. Gasser and H. Leutwyler, Ann. Phys. (N.Y.) 158
 (1984) 142; \np{\bf 250} (1985) 465.

\bibitem{smphen}D. Froidevaux, in {\em Proceedings of the
 ECFA Large Hadron Collider Workshop}, Aachen 1990, (G. Jarlskog and D.
 Rein, eds.), Vol. II, p. 444, and references therein.
\bibitem{gem}GEM Technical Design Report; GEM TN-93-262, SSCL-SR-1219;
 Submitted by the GEM Collaboration to the Superconducting Supercollider
 Laboratory (April 30, 1993).
\bibitem{rubbia}A. Rubbia, talk presented at the Conference
 {\em Beyond the Standard  Model IV}, Lake Tahoe,
 California, December 13-18, 1994.
\bibitem{ehlq}See, for example E.~J.~Eichten, C.~Quigg, I.~Hinchliffe
          and K.~D.~Lane, \rmp{\bf 56} (1984) 579.

\bibitem{kqq}V. Barger, J. L. Lopez and W.~Putikka, Int. J. Mod. Phys.
{\bf A3} (1988) 2181.
\bibitem{glover}E. W. N. Glover and J. J. van der Bij, \np{\bf 321}
 (1989) 561;\\
 U. Baur and E. W. N. Glover, in {\em Proceedings of the
 ECFA Large Hadron Collider Workshop}, Aachen 1990, (G.~Jarlskog and
  D.~Rein, eds.), Vol. II, p. 570;\\
 U. Baur and E. W. N. Glover, \prd{\bf 44} (1991) 99.
\bibitem{nisati}A. Nisati, in {\em Proceedings of the  ECFA Large Hadron
     Collider Workshop}, Aachen 1990, (G.~Jarlskog and   D.~Rein, eds.),
        Vol. II, p. 492.
\bibitem{CDF} CDF Collaboration, F. Abe et al.,
  Fermilab-Pub-94/0097-E.

\bibitem{djouadigl}A. Djouadi, M. Spira and P. Zerwas, \pl{\bf 264}
(1991) 440.
\bibitem{gr}D. Graudenz, M. Spira and P. Zerwas, \prl{\bf 70} (1993) 1372.
\bibitem{kz} Z. Kunszt and F. Zwirner, \np{\bf 385} (1992) 3.
\bibitem{effw}M. Chanowitz and M. K. Gaillard, \pl{\bf 142} (1984) 85;\\
  G. Kane, W. Repko and W. Rolnik, \pl{\bf 148} (1984) 367;\\
  S. Dawson, \np{\bf 249} (1985) 42.
\bibitem{chanowitz} M. Chanowitz and M. K. Gaillard, \np{\bf 261}
    (1985) 379.
\bibitem{repko}A. Abbasabadi and W. W. Repko, \prd{\bf 36} (1987) 289.
\bibitem{baur}U. Baur and E. W. N. Glover, \np{\bf 347} (1990) 12.
\bibitem{rogerio}J. Bagger, V. Barger, K. Cheung, J.~Gunion, T.~Han,
   G.~A.~Ladinsky, R.~Rosenfeld and C.-P.~Yuan, \prd{\bf 49} (1994) 1246.
\bibitem{kn:valencia92}G. Valencia and S.~D.~Willenbrock, \prd {\bf 46}
  (1992) 2247;\\
  S.~D.~Willenbrock and G. Valencia, \pl{\bf 247} (1990) 341.
\bibitem{kn:moriond} E. W. N. Glover, in {\em High Energy Hadronic
  Interactions '91}, Proceedings of the 26th Rencontre de Moriond, Les
  Arcs, Savoie, France, edited by J. Tran Thanh Van (Editions
  Frontieres, Gif-sur-Yvette, 1991), p.~161.

\bibitem{kn:seez}C. Seez et al., in {\em Proceedings of the
 ECFA Large Hadron Collider Workshop}, Aachen 1990, (G.~Jarlskog and
  D.~Rein, eds.), Vol. II, p. 474.
\bibitem{kn:wcorr}J. Fleischer and F. Jegerlehner, \prd{\bf 23} (1981)
  2001;\\
 N. Sakai, \prd{\bf 22} (1980) 2220;\\
 B. A. Kniehl, \np{\bf 357} (1991) 439;\\
 S.~G.~Gorishny, A. L. Kataev, S. A. Larin and L. R. Surguladze, Mod.
 Phys. Lett. {\bf A5} (1990) 2703;\\
 B. A. Kniehl, \pr{\bf 240} (1994) 211.

\bibitem{kn:kks} E. Braaten and J. Leveille, \prd{\bf 22} (1980) 715;\\
  M. Drees and K. Hikasa, \pl{\bf 240} (1990) 445; {\bf 262} (1991) 497;\\
  R. Kleiss, Z. Kunszt and W. J. Stirling, \pl{\bf 253} (1991) 269.
\bibitem{haber} H.~E.~Haber, preprint SCIPP 94/39, hep-ph 9501320.



\end{thebibliography}
\end{document}